\input harvmac
\input epsf

\overfullrule=0pt
\abovedisplayskip=12pt plus 3pt minus 1pt
\belowdisplayskip=12pt plus 3pt minus 1pt
%

\def\bar{\overline}

\def\bigone{\hbox{1\kern -.23em {\rm l}}}
\def\ZZ{\hbox{\zfont Z\kern-.4emZ}}
\def\half{{\litfont {1 \over 2}}}

\def\tr{{\rm tr}\,}
\def\Tr{{\rm Tr}\,}
\def\Pf{{\rm Pf}\,}
\def\hA{{\hat A}}
\def\hF{{\hat F}}

\def\cl{{\rm cl}}
\font\litfont=cmr6

\def\bx{{\bar x}}
\def\bQ{{\bar Q}}

\def\rmi{{\rm i}}

\lref\schomerus{V. Schomerus, {\it ``D-Branes and Deformation
Quantization''}, hep-th/9903205; JHEP {\bf 06} (1999) 030.}
\lref\myers{R. C. Myers, {\it ``Dielectric-Branes''}, hep-th/9910053;
JHEP {\bf 12} (1999) 022.}
\lref\ramswati{M. Van Raamsdonk and W. Taylor, {\it ``Multiple
Dp-branes in Weak Background Fields''}, hep-th/9910052;
Nucl.Phys. {\bf B573} (2000) 703.}
\lref\watirev{W. Taylor, {\it ``The M(atrix) Model of M Theory''},
hep-th/0002016.}
\lref\seiwit{N. Seiberg and E. Witten, {\it ``String Theory and
Noncommutative Geometry''}, hep-th/9908142; JHEP {\bf 09} (1999) 032.}
\lref\seibnew{N. Seiberg, {\it ``A Note on Background
Independence in Noncommutative Gauge Theories, Matrix Model and
Tachyon Condensation''}, hep-th/0008013.}
\lref\ishibashi{N. Ishibashi, {\it ``A Relation Between
Commutative and Noncommutative Descriptions of D-branes''},
hep-th/9909176.}
\lref\cornalba{L. Cornalba and R. Schiappa, {\it ``Matrix Theory Star
Products from the Born-Infeld Action''}, hep-th/9907211\semi
L. Cornalba, {\it ``D-brane Physics and
Noncommutative Yang-Mills Theory''}, hep-th/9909081.}
\lref\gmsone{R. Gopakumar, S. Minwalla and A. Strominger, 
{\it ``Noncommutative solitons''}, hep-th/0003160;
JHEP {\bf 05} (2000) 020.}
\lref\dmr{K. Dasgupta, S. Mukhi and G. Rajesh, {\it ``Noncommutative
Tachyons''}, hep-th/0005006; JHEP {\bf 06} (2000) 022.}
\lref\hklm{J. Harvey, P. Kraus, F. Larsen and E. Martinec, {\it
``D-Branes and Strings as Noncommutative Solitons''}, hep-th/0005031, 
JHEP {\bf 07} (2000) 042.}
\lref\hkl{J. Harvey, P. Kraus and F. Larsen, {\it ``Tensionless 
Branes and Discrete Gauge Symmetry''}, hep-th/0008064.}
\lref\sochichiu{C. Sochichiu,
{\it ``Noncommutative Tachyonic Solitons. Interaction with Gauge
Field''}, hep-th/0007217; JHEP {\bf 08} (2000) 026.}
\lref\gmstwo{R. Gopakumar, S. Minwalla and A. Strominger,
{\it ``Symmetry Restoration and Tachyon Condensation in Open String
Theory''}, hep-th/0007226.}
\lref\sennew{A. Sen, {\it ``Some Issues in Noncommutative Tachyon
Condensation''}, hep-th/0009038.}
\lref\sencs{A. Sen, {\it ``Supersymmetric World-Volume Action For
Non-BPS D-Branes''}, hep-th/9909062; JHEP {\bf 10} (1999) 008.}
\lref\hork{P. Ho\v rava, {\it ``Type IIA D-Branes, K-Theory and Matrix
Theory''}, hep-th/9812135; Adv. Theor. Math. Phys. {\bf 2} (1999)
1373.}
\lref\billo{M. Bill\`o, B. Craps and F. Roose, {\it ``Ramond-Ramond Coupling
of Non-BPS D-Branes''}, hep-th/9905157; JHEP {\bf 06} (1999) 033.}
\lref\kennedy{C. Kennedy and Wilkins, {\it ``Ramond-Ramond Couplings on
Brane-Anti-Brane Systems''}, hep-th/9905195; Phys. Lett. {\bf B464}
(1999) 206.}
\lref\twosens{B. Janssen and P. Meessen, {\it ``A Nonabelian
Chern-Simons Term for Non-BPS D-Branes''}, hep-th/0009025.}
\lref\garousi{M.R. Garousi, {\it ``Tachyon Couplings on Non-BPS
D-Branes and Dirac-Born-Infeld Action''}, hep-th/0003122;
Nucl.Phys. {\bf B584} (2000) 284.}
\lref\bergpanda{E.A. Bergshoeff, M. de Roo, T.C. de Wit, E. Eyras and 
S. Panda,  {\it ``T-Duality and Actions for Non-BPS D-Branes''}, 
hep-th/0003221; JHEP {\bf 05} (2000) 009.}
\lref\kluson{J. Kluso\v n, {\it ``D-Branes in Type IIA and Type IIB
Theories from Tachyon Condensation''}, hep-th/0001123.}
\lref\witnc{E. Witten, {\it ``Noncommutative Tachyons and String Field 
Theory''}, hep-th/0006071.}
\lref\mandalrey{G. Mandal and S.-J. Rey, {\it ``A Note on D-Branes of 
Odd Codimensions from Noncommutative Tachyons''}, hep-th/0008214.}
\lref\jmw{D. Jatkar, G. Mandal and S. Wadia, {\it ``Nielsen-Olesen
Vortices in Noncommutative Abelian Higgs Model''}, hep-th/0007078.}
\lref\hassan{S.F. Hassan and R. Minasian, {\it ``D-brane Couplings, 
RR Fields and Clifford Multiplication''}, hep-th/0008149.}
\lref\wyllard{N. Wyllard, {\it ``Derivative Corrections to D-Brane
Actions with Constant Background Fields''}, hep-th/0008125.} 

{\nopagenumbers
\Title{\vbox{
\hbox{hep-th/0009101}
\hbox{TIFR/TH/00-50}}}
{Chern-Simons Terms on Noncommutative Branes}
\centerline{{Sunil Mukhi}\footnote{}{E-mail: mukhi@tifr.res.in,
nemani@tifr.res.in} and {Nemani V. Suryanarayana}}
\vskip 8pt
\centerline{\it Tata Institute of Fundamental Research,}
\centerline{\it Homi Bhabha Rd, Mumbai 400 005, India}

\vskip 2truecm
\centerline{\bf ABSTRACT}
\medskip
We write down couplings of the fields on a single BPS Dp-brane with
noncommutative world-volume coordinates to the RR-forms in type II
theories, in a manifestly background independent way. This generalises
the usual Chern-Simons action for a commutative Dp-brane. We show that
the noncommutative Chern-Simons terms can be mapped to Myers terms on
a collection of infinitely many D-instantons. We also propose
Chern-Simons couplings for unstable non-BPS branes, and show that
condensation of noncommutative tachyons on these branes leads to the
correct Myers terms on the decay products.
\vfill
\Date{September 2000}
\eject}
\ftno=0

\listtoc
\writetoc

\newsec{Introduction and Review}

Much insight has been gained into the dynamics of brane decay using
noncommutativity. Turning on a constant NS-NS 2-form B-field along the
world-volume of a D-brane, one finds that the world-volume action
becomes a noncommutative field theory\refs{\schomerus,\seiwit}. On an
unstable, non-BPS D-brane, one gets a noncommutative field theory
involving tachyonic scalars, which generically possesses
noncommutative static soliton solutions\refs\gmsone\ at large values
of the noncommutativity parameter. These solutions can be
used\refs{\dmr,\hklm} to describe the decay of these
branes. Subsequently it was understood that large noncommutativity is
irrelevant to the problem, if one chooses the correct solution
including gauge field excitations along with the
tachyon\refs{\sochichiu,\gmstwo}. A beautiful background-independent
formulation of this, exhibiting the relation to matrix theory, was
given by Seiberg recently\refs\seibnew.

Many of the above works deal with bosonic D-branes, which have a DBI
action similar to that for the bosonic fields of D-branes in
superstring theory. However, a unique property of branes in
superstring theory is that they all have topological couplings of
Chern-Simons type to the Ramond-Ramond closed-string backgrounds. In
what follows, we will propose explicit expressions for the
Chern-Simons couplings on BPS D-branes in the presence of
noncommutativity.

We will also find analogous couplings on non-BPS branes, and show that
these terms provide a useful testing ground for the conjectures
involving brane decay via noncommutative solitons. One key property of
noncommutative solitons is that they can produce $N$ lower D-branes
starting from a single higher brane. In recent times it has emerged
that collections of $N$ D-branes have extra commutator couplings in
their world-volume theory, to all RR potentials, that do not appear
for a single D-brane\refs\myers. Hence one should expect to find such
terms starting with a noncommutative D-brane. We will see that such
terms indeed arise.

A (Euclidean) Dp-brane in bosonic string or superstring
theory has a Dirac-Born-Infeld action for massless modes that can be
written
\eqn\dbione{
S_{DBI} = T_p\int d^{p+1} x~ 
\sqrt{\det\big(g_{ij} + F_{ij} + B_{ij}\big)} }
where $i=1,\ldots,p+1$. $g_{ij}$ and $B_{ij}$ are pull-backs of the
spacetime metric and NS-NS $B$-field onto the brane world-volume,
$T_p={(2\pi)^{1-p\over2}\over g_s}$, and we have set
$2\pi\alpha'=1$. We are working in static gauge and in the DBI limit,
i.e. with constant fields.

The relevant results of ref.\refs\seiwit\ can be summarised as
follows. Let us assume for definiteness that we are dealing with a
Euclidean Dp-brane with an even-dimensional world-volume (so $p$ is
odd) and a constant non-zero background $B$-field has been turned on
over all $p+1$ directions. In the presence of this $B$-field, the
dynamics of the world-volume fields of the brane is equivalently
described by an action in terms of noncommutative variables given by
\eqn\dbitwo{
\hat{S}_{DBI} = \hat{T}_p\int d^{p+1} x~ 
\sqrt{\det\big(G_{ij} + \hat{F}_{ij} + \Phi_{ij}\big)}}
where $\hat{T}_p = {(2\pi)^{1-p\over2}\over G_s}$, and products of
fields appearing in this equation are understood to be $*$ products
with parameter $\theta$:
\eqn\starprod{
f(x) * g(x) \equiv e^{{i\over2}\theta^{ij}\del_i \del'_j} f(x)
g(x')\big|_{x=x'} }

The new field strength $\hat{F}$ is given in terms of the commutative
gauge field by the Seiberg-Witten transform. In the case of a single
Dp-brane with constant $\hat{F}$ (rank one commutative gauge field
$F$), $\hat{F}$ is related to $F$ by
\eqn\Ffrln{
F = \hat{F}{1\over {1-\theta \hat{F}}}}
The parameters $G_{ij}, \Phi_{ij}, G_s$ and $\theta^{ij}$ are given in 
terms of the commutative variables $g, B, g_s$ by
\eqn\params{
\eqalign{
{1\over G + \Phi} &= -\theta + {1\over g+B} \cr
G_s &= g_s \left(\det(G+\Phi)\over \det(g+B)\right)^\half }}
Because these equations determine four parameters in terms of three
physical (constant) backgrounds, there is a freedom in the
description. This can be seen either as the freedom to choose the
noncommutativity parameter $\theta$ at will (in which case $G_{ij},
\Phi_{ij}, G_s$ are determined), or the freedom to choose $\Phi_{ij}$, 
in which case $G_{ij}, \theta^{ij}, G_s$ are determined).

Three choices of the description are particularly interesting. For
$\Phi_{ij}=B_{ij}$, we have $\theta=0, G_s=g_s$ and
$G_{ij}=g_{ij}$. This is the commutative description. For $\Phi=0$ we
have some definite value of $\theta$ and the remaining parameters,
such that all the noncommutativity is absorbed into $\hF$ and the
$*$ product. And for $\Phi_{ij}=-B_{ij}$, we have the special values:
\eqn\specialval{
\theta^{ij} = (B^{-1})^{ij},\qquad
G_{ij} = - B_{ik}g^{kl}B_{lj},\qquad
G_s = g_s \sqrt{\det B\over \det g} }
This last choice has the special feature of manifest background
independence. This means the following: remaining within the
description via $\Phi=-B$, we can vary $B$, keeping $g,g_s$ and $F+B$
fixed. Then it was shown\refs\seiwit\ that the DBI action is invariant 
under this change. Let us rewrite the DBI action in a way that makes
this background-independence manifest.

Substituting $\Phi=-B=-\theta^{-1}$ into Eq.\dbitwo, we find that
in this description: 
\eqn\dbithree{
\eqalign{
\hat{S}_{DBI} &= \hat{T}_p \int d^{p+1} x~ 
\sqrt{\det\big(G_{ij} + \hat{F}_{ij} - \theta^{-1}_{ij}\big)}\cr
&= T_p \sqrt{\det g}\sqrt{\det \theta}
\int d^{p+1} x~ 
\sqrt{\det\big(-\theta^{-1}_{ik}g^{kl}\theta^{-1}_{lj} 
+ \hat{F}_{ij} - \theta^{-1}_{ij}\big)}\cr
&=T_p 
\int d^{p+1} x~ {\Pf Q\over \Pf \theta} 
\sqrt{\det\big(g_{ij} + (Q^{-1})_{ij}\big)}\cr}}
where $\Pf$ denotes the Pfaffian of an antisymmetric
matrix, and we have defined
\eqn\defq{
Q^{ij} = \theta^{ij} - \theta^{ik}\hF_{kl}\theta^{lj} }
Note that this differs by a sign from the corresponding definition in
Ref.\refs\seiwit. 

It turns out that $Q^{ij}$ is background-independent in the sense
explained above\refs\seiwit. This follows from Eqs.\Ffrln\ and
\specialval, which give:
\eqn\qback{
F+B = \hF{1\over 1-\theta \hF} + {1\over \theta} = {1\over Q}}

The final step is to note\refs\gmsone\ that the integral over
noncommutative variables $x^i$ satisfying
\eqn\noncommx{
[x^i,x^j] = i\theta^{ij} }
can be re-expressed in terms of a trace $\Tr$ over a Hilbert space of
operators whose commutation relations are independent of $\theta$. The
transcription is:
\eqn\transc{
\int d^{p+1}x \rightarrow \Tr (2\pi)^{p+1\over 2} \Pf\theta}
from which we finally get:
\eqn\dbifour{
\hat{S}_{DBI} = {2\pi\over g_s} 
\Tr \Pf Q \sqrt{\det\big(g_{ij} + (Q^{-1})_{ij}\big)}}
In this form, the background independence is manifest, from the
background-independence of $Q^{ij}$ and the fact that all the other
quantities in this action are closed-string quantities, which are
manifestly $B$-independent.

We can go one step further to obtain a useful insight into the meaning
of this action. Since $\hF$ and therefore $Q$ is constant, we can
try to find coordinates $X^i$ which satisfy
\eqn\noncommX{
[X^i,X^j]= iQ^{ij} }
Then we can use Eq.\transc\ in the reverse direction, with the
noncommutativity parameter being $Q$ instead of $\theta$:
\eqn\ncrev{
\Tr\rightarrow  {1\over (2\pi)^{p+1\over 2} \Pf Q}  \int d^{p+1}X }
This enables us to write the action Eq.\dbifour\ schematically as:
\eqn\dbifive{
\hat{S}_{DBI} = T_p
\int d^{p+1}X~ \sqrt{\det\big(g_{ij} + F_{ij} + B_{ij}\big)}}
where we have also used Eq.\qback. This deceptively simple expression
is just the {\it original} DBI action, in the original variables, but
with all products replaced by $*$ products with noncommutativity
parameter $Q={1\over F+B}$. The complication resides in the fact that
it has to be interpreted as an action for the dynamical variable
$\hA$, which is defined in terms of $F+B$ via Eqs.\qback\ and \defq.

Thus there is a ``short route'' from the commutative action
Eq.\dbione\ to the manifestly background-independent noncommutative
action Eq.\dbifive, bypassing the conversion to a formalism with
$\Phi$-dependence and the introduction of an ``open-string'' metric
and string coupling. It consists of introducing noncommuting
coordinates $X$ and interpreting their noncommutativity parameter as
the $Q$ defined in Eq.\defq. Alternatively, the prescription can be
defined to lead to the last line of Eq.\dbithree, namely insert a
factor of ${\Pf Q\over \Pf \theta}$ inside the integral, and replace
all products by $*$ products with parameter $\theta^{ij}$. In either
case, one must replace $(F+B)_{ij}$ wherever it occurs by 
$(Q^{-1})_{ij}$.

It is also straightforward to understand the relation between the new
$X$ satisfying Eq.\noncommX\ and the old ones satisfying
Eq.\noncommx. This relation is
\eqn\xandX{
X^i = x^i + \theta^{ij}\hA_j(x) }
It was discovered in Refs.\refs{\ishibashi,\cornalba} and used
recently in Ref.\refs\seibnew\ to argue a relation between matrix
theory and noncommutativity, exhibiting the role of matrix theory in
ensuring background-independence. For our purposes it is not necessary
to invoke matrix theory per se. The mere existence of the $X^i$
variables gives us a recipe to find a noncommutative DBI action
equivalent to the original commutative one, namely replacing $x^i$ by
$X^i$. 

In the following section we will apply this recipe to finding the
Chern-Simons terms for a noncommutative D-brane in type II superstring
theory.

\newsec{Chern-Simons Terms for a Noncommutative BPS D-brane}

It is convenient to start with a single Euclidean BPS D9-brane of type
IIB theory. We follow the normalization of Ramond-Ramond forms as in
Ref.\refs\myers. Let $\mu_p$ be the RR charge (which equals the
tension $T_p$ in these conventions) of a BPS Dp-brane. With these
conventions, the Chern-Simons terms on a D9-brane are:
\eqn\comcs{
S_{CS} = \mu_9 \int \sum_n C^{(n)}\, e^{B+F}}
where $C^{(n)}$ denotes the $n$-form RR potential. As is well-known,
the above expression involves the following prescription: the
exponential is to be expanded in a formal power series of wedge
products, and each term is then wedged with the appropriate RR form so
that the total form dimension is 10.

In the presence of non-zero constant NS-NS $B$-field background, we
have seen that the DBI action has a manifestly background-independent
description in which the world-volume becomes a noncommutative space
with noncommutativity parameter $Q={1\over F+B}$. Following the
procedure outlined in the previous section, we now write down the
Chern-Simons terms on the noncommutative D9-brane. Thus, in the
previous equation we simply replace $(F+B)_{ij}$, wherever it occurs,
by $(Q^{-1})_{ij}$, and insert a factor $\Pf Q\over
\Pf\theta$ under the integral sign. Then the noncommutative 
Chern-Simons term is:
\eqn\nccsx{
\hat{S}_{CS} = \mu_9 \int_x {\Pf Q\over \Pf\theta}
\sum_n C^{(n)}\, e^{\,Q^{-1}} }
where the underlying coordinates are the $x^i$ and all the products
are understood to be $*$ products defined in terms of $\theta$. Here
$Q^{-1}$ is understood to be the 2-form
$\half(Q^{-1})_{ij}\,dx^i\wedge dx^j$. Just as was done for the DBI
action, this can alternatively be expressed as the Hilbert-space
trace:
\eqn\nccshilb{
\hat{S}_{CS} = {2\pi\over g_s} \Tr \Pf Q~ \sum_n C^{(n)}\,
e^{\,Q^{-1}} }
This can also be re-expressed schematically in terms 
of ``original'' variables as:
\eqn\nccsX{
\hat{S}_{CS} = \mu_9 \int_X \sum_n C^{(n)}
e^{F+B} }
where this time the integral is over coordinates $X^i$ with
noncommutativity parameter $Q$.

The modification of the Chern-Simons terms on a D9-brane due to
noncommutativity can be understood physically as follows. Suppose
that, to start with, we turn on a constant $B$ field on the D9-brane
only along the directions $(x^9,x^{10})$. In noncompact space, this
effectively induces infinitely many D7-branes on the world-volume of
the D9-brane. But we know that such a collection of D7-branes must
couple to the RR 10-form potential via Myers terms\refs\myers, see
also Refs.\refs{\ramswati,\watirev}. Hence the expression we have
proposed in Eq.\nccshilb\ must contain these terms. Indeed, since we
have taken maximal noncommutativity there, our situation is more like
that of infinitely many D-instantons, which explains why the integral
has been completely replaced by a trace over an infinite dimensional
Hilbert space. This action then should contain Myers terms describing
the coupling of the D-instantons to all $p$-form potentials (even $p$)
in type IIB theory.

To see that this is so, let us first examine the term containing the
10-form $C^{(10)}$. The Myers term proportional to this form, on a
collection of $N$ D-instantons, looks like:
\eqn\myersten{
{2\pi\over g_s} \tr e^{i(\rmi_\Phi \rmi_\Phi)}C^{(10)} }
where $\rmi_\Phi$ is interpreted to mean the inner product of the
$N\times N$ matrix-valued scalar field $\Phi^i$ transverse to the
brane, with a lower index on the RR potential. Expanding the
exponential, we find the relevant Myers term is:
\eqn\expmyers{
\eqalign{
{2\pi\over g_s} \tr 
e^{i(\rmi_\Phi \rmi_\Phi)}C^{(10)}
&=
{2\pi\over g_s} \tr 
{i^5\over 5!}\,\Phi^{i_{10}}\Phi^{i_9}\ldots\Phi^{i_2}\Phi^{i_1}
\,C^{(10)}_{i_1 i_2\cdots i_9 i_{10}} \cr
&= -{2\pi\over g_s} \tr 
{1\over 5! 2^5}\,
\big(i[\Phi^{i_1},\Phi^{i_2}]\big)\ldots
\big(i[\Phi^{i_9},\Phi^{i_{10}}]\big) 
\,C^{(10)}_{i_1 i_2\cdots i_9 i_{10}} \cr } }
To compare, let us extract the 10-form term from Eq.\nccshilb:
\eqn\ncten{
\eqalign{
{2\pi\over g_s}\Tr (\Pf Q)\, C^{(10)} &=
{2\pi\over g_s}\Tr {1\over 5! 2^5}\, \epsilon_{i_1 i_2\cdots i_9 i_{10}}
Q^{i_1 i_2}\ldots Q^{i_9 i_{10}} 
{1\over 10!}\epsilon^{j_1 j_2\cdots j_9 j_{10}}
\,C^{(10)}_{j_1 j_2\cdots j_9 j_{10}}\cr
&= {2\pi\over g_s}\Tr {1\over 5! 2^5}\, 
Q^{i_1 i_2}\ldots Q^{i_9 i_{10}} 
\,C^{(10)}_{i_1 i_2\cdots i_9 i_{10}}\cr }}
Now in noncommutative theory, as we have seen, the identification with
matrix-valued transverse coordinates comes about as $Q^{ij}=
-i[X^i,X^j]$. Identifying the $X^i$ with $\Phi^i$, we see that the two 
expressions above agree.

Note that the above manipulations are covariant without specifying a
spacetime metric: $Q$ naturally has upper indices, while $C^{(10)}$ is
a 10-form and has lower indices. The $\epsilon$ symbols above have
constant components $\pm 1$, so they are tensor densities (the upper-
and lower-indexed ones having equal and opposite weight). Hence the
action is always a true scalar.

One can actually display the equivalence of the noncommutative action
Eq.\nccshilb\ to the Myers terms on a D-instanton in complete
generality. Like the commutative Chern-Simons terms in Eq.\comcs, the
noncommutative version Eq.\nccshilb\ also involves a prescription
whereby the exponential of the 2-form $Q^{-1}$ is expanded and its
wedge products taken with the appropriate RR form to make
10-forms. Since the integral has now been converted to a trace, these
10-forms are contracted with the totally antisymmetric
$\epsilon$-tensor in 10 dimensions (whose components are $\pm 1$) to
make 0-forms. By manipulations similar to those above, it is then easy
to derive the following identity:
\eqn\myersident{
\Tr \Pf Q~ \sum_n C^{(n)}
e^{\,Q^{-1}} = \Tr e^{-\half\rmi_Q} \sum_n C^{(n)} }
where $\rmi_Q$ acting on a 2-form $\omega_{ij}$ is defined as
$Q^{ji}\omega_{ij}$, and the prescription on the RHS is as follows:
the exponential is expanded out to such an order that the $n$-form on
which it acts is reduced by contractions to a scalar. This is
precisely Myers' prescription applied to D-instantons! Thus we see
that noncommutativity can be used to derive the Myers-Chern-Simons
terms, with the ``dotting'' prescription of Myers being dual to the
wedge prescription arising in conventional Chern-Simons terms on
branes. This in particular demonstrates that the expression
Eq.\nccshilb\ is nonsingular despite the appearance of $Q^{-1}$.

Let us next consider lower BPS D-branes, for example a (Euclidean)
D7-brane in type IIB. The noncommutativity parameter $\theta^{ij}$ is
now chosen to be maximal with respect to the eight Euclidean
directions $x^1,x^2,\ldots,x^8$. Therefore this D7-brane is T-dual to
a D9-brane with noncommutativity only along the first eight
directions. Note that on this D7-brane we now have transverse
coordinates $\Phi^9,\Phi^{10}$ that are functions of the noncommuting
brane world-volume coordinates $x^i,i=1,\ldots,8$. It follows that
$\Phi^9$ and $\Phi^{10}$ do not commute, or alternatively that they
are multiplied using the $*$ product. This suggests that in addition
to the Q-dependent modification to the Chern-Simons term as in
Eq.\nccsx, we also have Myers-type terms involving $\Phi^9,\Phi^{10}$.

Thus we claim that the Chern-Simons term on the D7-brane is:
\eqn\csseven{
\hat{S}_{CS} = \mu_7 \int_x {\Pf Q\over \Pf\theta}\, 
P\left[ e^{i(\rmi_\Phi * \rmi_\Phi)} \sum_n C^{(n)}\right] e^{\,Q^{-1}} }
where $P$ represents the pull-back of the transverse brane coordinates
to the world-volume (this pullback was not required while studying the
9-brane, since there were no directions transverse to the brane in
that case). Note that here, $Q$ is an $8\times 8$ rather than
$10\times 10$ matrix.

This expression is basically equivalent to Eq.\nccsx. To see this,
notice that the exponential term $e^{i(\rmi_\Phi * \rmi_\Phi)}$ is
equivalent to $1+ i(\rmi_\Phi * \rmi_\Phi)$ since higher powers cannot
contribute by antisymmetry of the object that they contract.  If we
convert the 8-dimensional integral above into a Hilbert-space trace,
and define $Q^{9,10}= -i[\Phi^9,\Phi^{10}]$, then it is not hard to
see that Eq.\csseven\ becomes equivalent to Eq.\nccsx. For example, in
Eq.\csseven, the term $(Q^{-1})_{9,10}$ is missing, but the term in
Eq.\nccsx\ where it would contribute by cancelling a factor of
$Q^{9,10}$ from the Pfaffian, arises instead from the 1 in $1+
i(\rmi_\Phi * \rmi_\Phi)$.

Thus we see that Chern-Simons terms on BPS D-branes in the presence of
noncommutativity are summarised by Eq.\nccsx, though for different $p$
they are more appropriately written in terms of a hybrid of $\Pf Q$
for the directions within the brane, and exponential terms of the
Myers type for the directions transverse to the brane, as in
Eq.\csseven\ for the case of the D7-brane.

\newsec{Chern-Simons Terms for a Noncommutative Non-BPS D-brane}

We now turn to the discussion of unstable, non-BPS D-branes and
determine the Chern-Simons action on their world-volumes in the
presence of noncommutativity. Type II superstring theories have
unstable D$p$-branes, where $p$ is odd in type IIA and even in type
IIB. Like the BPS branes, these branes too have Chern-Simons terms on
their world-volumes\refs{\sencs,\hork,\billo,\kennedy,\twosens} which
play an important role in arguing that their decay products (in the
commutative case) carry the right RR charges. However, these
Chern-Simons terms now depend on the tachyon field as well.

For a single unstable commutative $p$-brane, the CS term has been
argued to be:
\eqn\commcs{
S_{CS} = {\mu_{p-2}\over 2T_{min}}\int 
dT\, C^{(n)}\, e^{F+B}}
where $T_{min}$ is the value of the tachyon at the minimum of the
potential. Note that $dT$ is a 1-form, and the prescription for the
Chern-Simons term involves expanding the exponential so that, after
wedging with an appropriate RR potential {\it and} the tachyon factor
$dT$, we get a 10-form.

For the noncommutative case, let us work with a Euclidean D9-brane in
type IIA with noncommutativity over all the 10 directions.  Following
the procedure described in the previous sections, we can try to write
down the Chern-Simons action. However, there is an important subtlety
to be taken care of. The exterior derivative $dT$ of the tachyon in
the above equation must be promoted to a covariant derivative in the
noncommutative case. The covariant derivative in our notation is $D_i
=
\theta^{-1}_{ij}X^j$ with $X^i$ defined in Eq.\xandX. The problem is
that since $X^i$ is background-independent, $D_i$ depends explicitly
on $\theta$. We need to find a 1-form in place of $dT$ that is linear
in $[X^i,T]$ and also background-independent. The unique candidate
seems to be 
\eqn\covderiv{
{\cal D}_i T = -i\, (Q^{-1})_{ij}[X^j,T] }
where $Q$ has been defined in Eq.\defq. One piece of encouragement
for this replacement, besides background-independence, comes from the
fact that at $\hF=0$, $Q$ is the same as $\theta$, so our covariant
derivative reduces to the standard one in that case. We will see
another justification for it in the following section.

Accordingly, we propose the following Chern-Simons action for an
unstable D9-brane:
\eqn\csnbps{
\hat{S}_{CS} = {\mu_{8}\over 2T_{min}}\int_x {\Pf Q\over \Pf\theta}\,
{\cal D}T\, C^{(n)}\,e^{\,Q^{-1}}}
with ${\cal D}T$ as defined above. We will see in the following
section that this modification is crucial in reproducing expected
results after brane decay via noncommutative solitons.

Note that even the commutative Chern-Simons term of Eq.\commcs\ could
have higher-order corrections containing powers of $T$, as noted in
Refs.\refs{\kluson,\twosens}. In this case, the noncommutative version 
will also generalise in an obvious way.

\newsec{Noncommutative Solitons, Brane Decay and Myers Terms}

Now we can use the known classical solutions for noncommutative
tachyons representing lower dimensional unstable D-branes, and study
the Chern-Simons action expanded around these solutions.

Let us first review the relevant information in Ref.\refs\gmstwo,
which extended the observations in Refs.\refs{\dmr,\hklm}, about
noncommutative soliton solutions on unstable D-branes. However, unlike
these references, we work in the description $\Phi_{ij} = -B_{ij}$
following Seiberg\refs{\seibnew}.  In this description the action is
manifestly independent of both $\theta$ and $B$. Written as a trace
over a Hilbert space, the DBI action for an unstable D9-brane is:
\eqn\usdbitwo{
\hat{S}_{DBI} =  {2\pi \over g_s} {\rm Tr}\left[
V(T)\sqrt{\det\big(\delta_i^j - ig_{ik}[X^k, X^j]\big)} - f(T)\, 
g_{kl}\,[X^k, T][X^l, T] + \cdots\right]}
where $V(T)$ is the tachyon potential, and all products are understood
to be $*$ products. $f(T)$ is some function whose form we will not
need, although from Refs.\refs{\garousi,\bergpanda} it follows that it
is proportional to $V(T)$.

The above action can be obtained following the ``long route''
described in Section 1. However, it can also be obtained via the short 
route if we assume the formula in Eq.\covderiv. For this, we start
with the commutative action in the form\refs{\garousi,\bergpanda}:
\eqn\dbionenon{
S_{DBI} = T_p\int d^{p+1} x~ 
\sqrt{\det\big(g_{ij} + F_{ij} + B_{ij} + \del_i T \del_j T\big)} }
Using the prescription in Section 1 along with the one in
Eq.\covderiv, the noncommutative action written as a trace is:
\eqn\dbifournon{
\hat{S}_{DBI} = {2\pi\over g_s} 
\Tr \Pf Q \sqrt{\det\big(g_{ij} + (Q^{-1})_{ij} + {\cal D}_i T {\cal
D}_j T^\dagger \big)}}
Taking the Pfaffian factor inside the square root and expanding, we
easily recover Eq.\usdbitwo, with $f(T)=V(T)$\foot{It is important to note
that the expansion of the DBI action is being carried out for $|g|,
|\del T|^2 \ll |B+F|$, while in the commutative theory the DBI action
is expanded in the opposite regime, where $|\del T|^2, |B+F| \ll
|g|$.}. This provides additional confirmation of Eq.\covderiv.

Now one can extremise this action and find the classical equations of
motion for both the operators $T$ and $X^i$. For the equation of
motion of the tachyon field we get:
\eqn\teom{
V'(T)\sqrt{\det M} - f'(T)\,g_{kl}\,[X^k,T][X^l, T] + 2g_{kl}\,[X^k,
[X^l, T]\, f(T)] = 0}
and for the $X^i$ we get:
\eqn\xeom{
2g_{ij}\left[T, [T,X^j]\,f(T)\right] + i g_{kl}\left[X^l, V(T)\sqrt{det
M}(M^{-1})^k_i\right] = 0}
where we have defined:
\eqn\mdef{
(M)_i^j = \delta_i^j - i g_{ik}\,[X^k, X^j]}
The equations of motion above actually suffer from an ordering
problem, as does the DBI action itself. However, as was essentially
noted in Ref.\gmstwo, the classical solutions representing
noncommutative solitons turn out to be independent of any
rearragements of terms one might make above (so long as one does not,
of course, open out commutators).

In these variables we can find solutions for these equations that
correspond to the ``nothing state'' and to various lower dimensional
branes. As was pointed out by various authors\refs{\gmstwo,\seibnew},
each such state has in general a multitude of solutions. Recently
Sen\refs\sennew\ has argued that this apparent degeneracy of solutions
arises because the variables in which the DBI action is written are
not the correct variables at the end-point of tachyon
condensation. One has to use some combinations of these variables to
get rid of the unwanted degeracy of solutions. For our purpose,
however, it is enough to carry out the analysis using any one of the
many physically equivalent solutions.

Hence we take the following solution as the ``nothing state'':
\eqn\nstate{
\eqalign{
T_\cl &= T_{min} 1 \cr
X^i_\cl &= 0 \qquad {\rm for}\qquad i= 1,2,\dots,10}}
where 1 represents the identity operator.  For a codimension-two soliton
representing a $D7$ brane (say along the $x^1, x^2,\ldots,x^8$
directions), we choose the following solution:
\eqn\dseven{
\eqalign{
T_\cl &= T_{max} P_N + T_{min}(1-P_N)\cr
X^i_\cl &= P_N x^i\quad {\rm for}\quad i= 1,2,\ldots,8\cr
X^i_\cl &= 0 \quad {\rm for}\quad i= 9,10\cr}}
where $T_{max}$ and $T_{min}$ are the values of the tachyon field at
the extrema of $V(T)$, and $P_N$ is the level-$N$ projection operator
in the harmonic oscillator Hilbert space made out of the
noncommutative directions $x^8,x^9$. 

It is easy to see that for the ``nothing state'' solution Eq.\nstate,
the action vanishes identically. For the codimension-two soliton, using:
\eqn\tacpot{
V\Big(T_{max} P_N + T_{min}(1-P_N)\Big) = V(T_{max})\,P_N}
we find that the action is given by:
\eqn\actntwo{
\eqalign{\hat{S} &= {2\pi \over g_s}\Tr V(T_{max})P_N 
\sqrt{\det\big({\delta_i}^j + P_N g_{ik}\theta^{kj}\big)}\cr
&= {2\pi \over g_s}\,N 
\sqrt{\det\big({\delta_i}^j + g_{ik}\theta^{kj}\big)} }}
This expression is identified with the action for $N$ unstable $D7$
branes with all fluctuations set to zero (in particular, this means
that $T=T_{max}$ and ${\hat A}=0$).

The apparent $\theta$-dependence of this result is understood as
follows. We have obtained the action for unstable D7-branes in the
description with $\Phi=-\theta^{-1}$, and at $\hF=0$. For general
$\Phi$ the answer should be proportional to $\sqrt{\det(G+\Phi)}$.
Hence in the background-independent description, with
$\Phi=-\theta^{-1}$, it should be proportional to
$\sqrt{\det(G-\theta^{-1})}$, which is the case. Another way to put it
is that the $\theta$ in Eq.\actntwo\ is really $Q$, as defined in
Eq.\defq, and evaluated at $\hF=0$.

We could in fact have chosen the classical solution for $X^i$ to be
given by $X^i_\cl=P_N\bx^i$ for arbitrary $\bx^i$ satisfying
$[\bx^i,\bx^j]=i\bQ^{ij}$, this would have reproduced D7-branes in the
state with $Q^{ij} = \bQ^{ij}$. But the state with $\hA=0$ (hence
$\bQ^{ij}=\theta^{ij}$) is special in that it describes the final
D7-branes in their undecayed state.

We can now insert these solutions into the Chern-Simons term of
Eq.\csnbps\ and find the Chern-Simons action for the fluctuations of
these solitons. Here we will just demonstrate how to get Chern-Simons
terms for an unstable D7 brane of type IIA. For this, let us first
consider a specific term from Eq.\csnbps\ for an unstable D9 brane:
\eqn\csnbps{
\hat{S}_{CS} = {\mu_{8}\over 2T_{min}}\int_x {\Pf Q\over \Pf\theta}\,
\,(-i)(Q^{-1})_{ij}[X^j,T]\, C^{(9)}}
Now we condense the noncommutative tachyon as a codimension two
soliton given in Eq.\dseven\ to obtain $N$ D7-branes along $x^1,
x^2,...,x^8$ directions. We substitute 
\eqn\fltns{
T = T_\cl + \delta T, \qquad
X^i = X^i_\cl + \delta X^i}
with the classical solutions defined in Eq.\dseven, into
Eq.\csnbps. In \fltns\ we take the fluctuations $\delta T$ and $\delta
X^i$ to be independent of the $x^9, x^{10}$ directions. Then this CS
action will have the form of an action on a set of D7-branes. Indeed,
we saw in Eq.\actntwo\ that expanding the DBI action brings out a
projection operator $P_N$ for the Hilbert space along the
noncommutative directions $9,10$, so that the remaining $(\infty -N)$
modes become nonpropagating. The same projection arises in the CS
action because of the $P_N$ in the classical solution for the $X^i$.

Notice that with this classical solution we have $[X^i_\cl,T_\cl]=0$,
so the Chern-Simons action of the classical solution vanishes
identically. The decay product therefore has no ``winding brane
charge'', contrary to the claim in Refs.\refs{\dmr,\witnc}. This is
particularly reassuring in view of a recent argument\refs\hkl\ that some
of the configurations that were thought to carry such charge are pure
gauge under a certain discrete symmetry.

Now let us consider the case in which the 9-form $C^{(9)}_{i_1
i_2..i_9}$ has its indices along the directions $2,3,...,10$. In this
case the index $i$ in the 1-form $(Q^{-1})_{ij}[X^j,T]$ in Eq.\csnbps\
will have to be along the direction 1. In a coordinate basis in which
the matrix $Q^{ij}$ is in the canonical form, it follows that the
index $j$ must be 2. Then we end up with the following action for the
fluctuations $\delta T$ and $\delta X^i$:
\eqn\dsevenone{
\hat{S}^{UD7}_{CS} = {\mu_{6}\over 2T_{min}}{\rm tr}_N 
\int_x (-i)\,[\delta X^9, \delta X^{10}]\,(-i)(Q_\cl^{-1})_{12}
\,[X^2_\cl,\delta T]\, C^{(9)}_{2 3 \ldots 10}}
where $Q^{ij}_\cl=\theta^{ij}$. Now in the above expression,
$X^9,X^{10}$ are $N\times N$ matrices (because of the $P_N$
projection) and the Hilbert-space trace over the directions of the
noncommutative soliton becomes a trace over these matrices. Hence,
what we have found here is a Myers term on N unstable D7-branes.

Actually, the commutator $[\delta X^9, \delta X^{10}]$ that enters
above does not vanish even for $N=1$. This is because the $X^i$'s
appearing in it are not only $N\times N$ matrices, but also functions
of the remaining 8 coordinates which are multiplied using the $*$
product. This is due to our choice of noncommutativity over all the
directions, and not just over the two directions along the
noncommutative soliton solution and transverse to the final D7-brane.

Next we may consider the term where the indices on $C^{(9)}$ are
$1,2,\ldots,8,10$. In this case, $i$ in Eq.\csnbps\ must be 9 and
therefore $j$ is 10. The resulting term is:
\eqn\dseventwo{
\eqalign{
\hat{S}^{UD7}_{CS} &= {\mu_{6}\over 2T_{min}}{\rm tr}_N
\int_x (-i)\,[\delta X^9, \delta X^{10}]\,(-i)(\delta
Q^{-1})_{9,10}\,[\delta X^{10},\delta T]\, C^{(9)}_{12\ldots 8,10}\cr
&= {\mu_{6}\over 2T_{min}}{\rm tr}_N\int_x 
(-i)[\delta X^{10},\delta T]\, C^{(9)}_{12\ldots 8,10}\cr}}
where $\delta Q^{9,10}$ is shorthand for $-i[\delta X^9,\delta
X^{10}]$. This is a new kind of Myers term on $N$ unstable branes,
that has no analogue for BPS branes. It was discovered very recently
in Ref.\refs\twosens.

We see that the Chern-Simons terms on a noncommutative unstable
D-brane, Eq.\csnbps, beautifully reproduce the structure of extra
commutator terms that are expected to be present for an assembly of
$N$ unstable D-branes, lending further support to the idea that
noncommutative tachyons really do reproduce $N$ unstable D-branes of
codimension 2 and that the $U(N)$ of these D-branes is naturally
embedded in the $U(\infty)$ associated to noncommutativity.

\newsec{Discussion and Conclusions}

We have found Chern-Simons actions for D-branes of type IIB
superstring theory, with noncommutativity (a $B$-field) along their
world-volume. These actions are manifestly background-independent, and
also manifestly gauge-invariant as they are expressed in terms of
traces over a Hilbert space with $U(\infty)$ symmetry. The
noncommutative expressions are elegant and turn out to contain all
information about Myers terms on multiple branes.

In our work we focussed on Euclidean branes with maximal
noncommutativity, hence branes of even world-volume dimension. This
means that we studied BPS branes in type IIB, and unstable branes in
type IIA. The other cases: BPS branes in type IIA and unstable branes
in type IIB, can be studied by remaining in Minkoswki signature and
turning on a B-field over all the spatial directions. In this case,
the noncommutative actions will resemble actions for D0-branes rather
than D-instantons.

The noncommutative Chern-Simons terms found above are valid, as for
the Myers terms, in the static gauge for a D-brane with very slowly
varying fields. A covariant generalisation of this should be expected
to exist and could perhaps be found along the lines of
Ref.\refs\hassan. Extension to higher-derivative terms in the field
strength is also an interesting open question, for example one could
try to generalise the results of Ref.\refs\wyllard\ to the
noncommutative case.

It would also be interesting to check how the noncommutative CS terms
transform under T-duality --- they should form a consistent
collection, as for the case of the DBI
action\refs{\garousi,\bergpanda} and Chern-Simons
action\refs{\myers,\twosens} on single or multiple branes. One can
also hope to check that these terms, on unstable branes, give correct
results when evaluated on codimension-one noncommutative
solitons\refs\mandalrey. And, with our methods it should be
straightforward to write down the Chern-Simons terms on a
noncommutative brane-antibrane pair, generalising the result of
Ref.\refs\kennedy, and study noncommutative solitons on these
pairs\refs{\hklm,\witnc,\jmw}.
\bigskip

\noindent{\bf Acknowledgements:}
\medskip

We would like to thank Atish Dabholkar, Sumit Das, Sudhakar Panda,
Ashoke Sen, Sandip Trivedi and Spenta Wadia for useful discussions,
and Shanta de Alwis, Fawad Hassan and Edward Witten for helpful
correspondence.

\listrefs

\end